\documentclass[aps,amsmath,showpacs,amsfonts,10pt]{revtex4}
\usepackage{epsfig,graphicx}
\usepackage[english]{babel}
\usepackage{amsfonts}
\usepackage{amsmath}
\usepackage{latexsym}
\usepackage{graphics,bm}
\usepackage{dcolumn}
\usepackage{bm}
\usepackage{rotating}

\begin{document}

\title{The Quantum Acousto-Optic Effect in Bose-Einstein Condensate}

\author{Aranya B Bhattacherjee}
\affiliation{Department of Physics, ARSD College, University of Delhi(South Campus), New Delhi-110021, India }

\begin{abstract}
We investigate the interaction between a single mode light field and an elongated cigar shaped Bose-Einstein condensate (BEC), subject to a temporal modulation of the trap frequency in the tight confinement direction. Under appropriate conditions, the longitudinal sound like waves (Faraday waves) in the direction of weak confinement acts as a dynamic diffraction grating for the incident light field analogous to the acousto-optic effect in classical optics. The change in the refractive index due to the periodic modulation of the BEC density is responsible for the acousto-optic effect. The dynamics is characterised by Bragg scattering of light fom the matter wave Faraday grating and simultaneous Bragg scattering of the condensate atoms from the optical grating formed due to the interference between the incident light and the diffracted light fields. Varying the intensity of the incident laser beam we observe the transition from the acousto-optic effect regime to the atomic Bragg scattering regime, where Rabi oscillations between two momentum levels of the atoms are observed. We show that the acousto-optic effect is reduced as the atomic interaction is increased.
\end{abstract}


\maketitle

\section{Introduction}

When a liquid traversed by compression waves of shorter wavelength is irradiated by visible light, diffraction phenomenon similar to that due to a grating is produced. This phenomenon was predicted by Brillouin \cite{Brillouin21} in 1921 and subsequently verified experimentally by Debye et al. \cite{Debye32}, and Lucas et al. \cite{Lucas32}. This effect is considered to be an important discovery since this led to the interesting application of acoustic modulation of optical radiation by naturally birefringent crystal. The linear acousto-optic (also known as photo-elastic) effect involves the first order changes in the optical properties of insulators due to acoustic strain. The acousto-optic effect, however has not yet been suggested for the most recently created state of matter, the Bose-Einstein condensate (BEC).

An important question in this context is the generation of compression waves in a BEC which can act as a dynamic diffraction grating for the incident light field. The generation of such a kind of longitudinal sound like waves was first predicted by Staliunas et al. \cite{Stalinus02, Stalinus04} and experimentally achieved recently in an elongated BEC in the direction of weak confinement by periodically modulating the radial trap frequency \cite{Engels07}. The waves thus generated are known as Faraday waves. The modulation of the radial trap frequency leads to a periodic change of the density of the cloud in time, which is equivalent to a change of the nonlinear two-body interaction. This in turn, lead to the parametric excitation of longitudinal soundlike waves in the direction of weak confinement. In this work, we report on the diffraction of light waves from a BEC subjected to a temporal modulation of the trap frequency in the tight confinement direction. In contrast to the classical acousto-optic effect in liquids and crystals, the two-body atom-atom interactions and the back action of the light fields on the atomic medium in the present system generates new and interesting physics. The light fields create an effective optical grating inside the BEC which Bragg scatters the very atoms which are generating the matter-wave grating for the light field. This complex interplay between the atoms and the photons gives the acousto-optic effect a quantum character unseen in the classical case. The atomic interactions are tunable by Feshbach resonances and this allows us also to investigate the effects of the mean-field atomic interaction on the diffraction process.

In the context of interaction of light fields with BEC, the long coherence time of the BEC offers the possibility to study collective optical effects of atoms in atom optics such as Bragg scattering of atoms from standing wave laser field and superradiant Rayleigh scattering. The photon exchange between different ultracold atoms result in a long-range inter-atom interaction which under appropriate conditions leads to an effective "Kerr-type" nonlinearity of atomic waves \cite{Zhang}. Analogous to nonlinear quantum optics, multiwave mixing has been predicted \cite{Goldstein} and observed \cite{Ketterle, Phillips} in BEC. A nonlinear quantum optics approach to the theory of optically driven BEC's was given by \cite{Moore1}, where it was shown that quantum noise can generate entangled atom-photon pair which exhibit nonclassical correlations similar to those seen between photons in optical parametric amplifier. The recent observation of superradiant Rayleigh scattering from BEC's \cite{Inouye, Schneble} has significantly extended our knowledge about collective emission processes. The basic theory of superradiant Rayleigh scattering has been discussed by a number of authors \cite{Moore}-\cite{Zobay05}.

\section{Faraday Pattern as a Dynamic Diffraction Grating}

 In our theoretical treatment, we consider an elongated cigar shaped condensate consisting of $N$ two-level atoms, oriented along the $z$ axis. The trapping potential of the ground state is periodically modulated and is written as $V_{1}(\vec{r},t)=\dfrac{m}{2}\left[ \Omega_{\bot}^{2} (t)\left( x^2+y^2\right)+\Omega_{||}^{2} z^2 \right]$, with $\Omega_{\bot}=\bar \Omega_{\bot}\left( 1+\alpha \cos {\Omega t}\right) $. Here $\Omega_{\bot}$ and $\Omega_{||}$ are the trap frequencies along the radial and the longitudinal($z$) direction respectively. Also, $\bar \Omega_{\bot}$ is the amplitude of the radial trap frequency and $\Omega$ is the frequency of the trap modulation. The condition $\Omega_{\bot}$ $>>$ $\Omega_{||}$ is assumed. Due to the periodic modulation of the trap, the ground state wavefunction is also modulated as:

\begin{equation}
\psi_{1}(\vec r, t)= \psi_{0}(\vec r ,t) \left\lbrace 1+w(t) \cos{\vec k_{s}. \vec r}\right\rbrace . 
\end{equation}

\begin{figure}[t]
\hspace{2.5cm}
\includegraphics{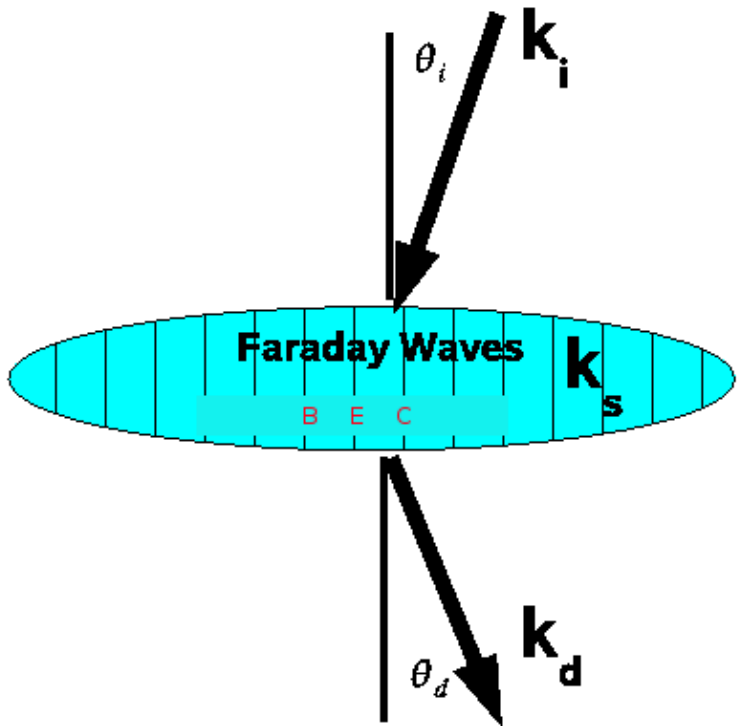} 
\caption{Schematic diagram showing the Bragg scattering of light incident at an angle $\theta_{i}$ with wavevector $\vec{k_{i}}$ by Faraday waves of wavevector $\vec{k_{s}}$ in an elongated cigar shaped BEC. The incident light is diffracted at an angle $\theta_{d}$ and a wavevector $\vec{k_{d}}$. The polarization of the incident light is chosen to be in the plane defined by the $z$ axis and the direction of the wave-vector of the incident light so that the superradiant Rayleigh scattering is suppressed. }
\label{figure1}
\end{figure}

Here, $\psi_{0}(\vec r ,t)$ is the wavefunction of the homogeneous BEC in the absence of any modulations and $w(t)$ is the complex valued amplitude of the perturbation. The periodically modulated BEC is exposed to a far off-resonant laser field whose polarization is parallel to the longitudinal axis ($z$). Under proper conditions, part of the input optical beam is diffracted into a new direction while simultaneously being shifted in frequency and wavenumber by an amount equal to the frequency of the Faraday wave ($\omega_{s}$) and wavenumber ($k_{s}$) respectively. Unlike the classical acousto-optic effect, here the choice of the polarization of the incident light is extremely important since the dominant competing process superradiant Rayleigh scattering can completely mask the acouto-optic effect in a BEC.  According to ref. \cite{Inouye}, the growth rate of $N_{j}$ recoiling atoms is $\dfrac{dN_{j}}{dt}=R N_{0} \dfrac{sin^{2} \theta_{j}}{8 \pi/3}\Omega_{j}(N_{j}+1)$, where $R$ is the rate for single-atom Rayleigh scattering, which is proportional to the laser intensity, $N_{0}$ is the initial number of atoms, $\theta_{j}$ is the angle between the polarization of the incident light and the direction of emission ($z$ axis) and $\Omega_{j}$ is the solid angle.  For polarization parallel to the $z$ axis ($\theta_{j}=0$), light emission into the end-fire mode is suppressed and the higher order momentum side modes of the condensate are not generated and the distribution of the atoms follow the dipolar pattern of normal Rayleigh scattering. For perpendicular polarization ($\theta_{j}=\pi/2$), photons are emitted along the $z$ axis and the recoiling atoms appear as highly directional beams at an angle $45^{0}$ with respect to the $z$ axiz. This is the superradiant Rayleigh scattering which can destroy the acousto-optic effect which we wish to study in this work. Due to the highly directional nature of the momentum transfer in the superradiant scattering event, the kinetic energy gained by the atoms due to the recoil , far exceeds the kinetic energy gained due to the periodic modulation of the trap. In the case of normal Rayleigh scattering, due to the random nature of the spontaneous emission, the direction of the recoil is random, leading to momentum diffusion. The amount of atoms lost from the original condensate in the superradiant Rayleigh scattering is extremely high compared to that from normal Rayleigh scattering\cite{Inouye}. In order to suppress super-radiant Rayleigh scattering, the incident light beam is taken to be linearly polarized in the plane defined by the condensate axis and the wave vector of the light \cite{Inouye,Kurn}. Note that experimentally for parallel polarization and small detuning, yet another new effect is produced namely the superradiant Raman scattering \cite{Schneible, Yukata}. In the superradiant Raman scattering, only two modes of the recoiling condensate atoms are populated and the decoherence is also much faster compared to superradiant Rayleigh scattering due to inhomogeniety in the magnetic field. This fast decoherence rate can be traced to a new type of coherence grating that is formed and this grating is highly sensitive to magnetic fields. A slightly inhomogeneous magnetic field could be used to suppress the coherence grating which could also interfer with the quantum acousto-optic effect. There is yet another effect which is in phase with the acousto-optic effect, the Bragg scattering of the condensate atoms from the optical lattice created due to the interference between the $z$ component of the incident and the diffracted light waves \cite{Martin}. In section IIB, we treat the light field classically with fixed intensity and treat the atoms in the mean field regime. This treatment is useful to gain insight into the dynamics of the BEC density and in particular the time evolution of the shape of the cloud. In section III, we take the light as quantum and this treatment helps us to study the time evolution of the light fields. We will show in section III that the Bragg scattering of the atoms can be minimized if the decay rate of the light fields are greater than the atom-field coupling strength (this can be controlled by the intensity of the light field).

\subsection{Bragg scattering condition}

Figure 1, shows the schematic setup with which one can observe the acousto-optic effect in an elongated cigar shaped BEC. The Faraday waves act as dynamic diffraction grating and Bragg scatter the incident light. Let the incident and the diffracted optical field consist of two plane waves with frequencies $\omega_{i}$ and $\omega_{d}$ respectively. The two waves propagate along $\vec k_{i}$ and $\vec k_{d,\ell}$, respectively, where $\vec k_{i}$ in general is not parallel to $\vec k_{d,\ell}$. There are two main diffraction regimes, the Raman-Nath regime and the Bragg regime \cite{Born}. When the parameter $\chi=\epsilon k_{L}/k_{s}\sin{\theta_{i}}>1$, we are in the Raman-Nath regime which is characterized by the presence of many diffraction orders and when $\chi<1$, we are in the Bragg regime which is characterized by the presence of only the first order diffraction. Here $\epsilon$ and $k_{L}$ are the dielectric constant of the BEC and the wavenumber of the laser beam. Energy and momentum conservation leads to the following relations:

\begin{equation}
\omega_{d,\ell}-\omega_{i} = \ell \omega_{s}
\label{sign1}
\end{equation}

\begin{equation}
\vec k_{d,\ell}-\vec k_{i} =\ell \vec k_{s}
\label{sign2}
\end{equation}

Here $\ell=0,\pm1,\pm2...$ is the order of diffraction and $\vec k_{d,\ell}$ is the diffracted wavevector of the $\ell^{th}$ order. The interaction can be viewed in the following terms. A photon with energy $\hbar \omega_{i}$ and momentum $\hbar \vec k_{i}$ is incident on a Faraday wave of frequency $\omega_{s}$ and wave momentum $\hbar \vec k_{s}$. The incident photon and a phonon of the Faraday wave are annihilated, giving rise instead to a new photon at $\omega_{d,\ell}$ , $\vec k_{d,\ell}$. 

The Eqn. \ref{sign2} yields the usual familiar diffraction condition for $k_{i}=k_{d,\ell}=k_{L}$:

\begin{equation}
k_{L} (\sin{\theta_{d,\ell}-\sin{\theta_{i}}})=\ell k_{s}
\label{Braggcond1}
\end{equation}

The periodic nature of the Faraday waves play a role similar to that of a regular arrangement of atomic planes. In Eq.(4), we have assumed $k_{i} = k_{d}=k_{L}$. Also $k_{L} \approx 10^6 m^{-1}$ and $k_{s} \approx 10^5 m^{-1}$.

\subsection{BEC dynamics in the presence of classical light field and Faraday instability}

The system consisting of the periodically modulated condensate with $N$ two-level atoms and the light fields is described by the Hamiltonian

\begin{eqnarray}
H&=&\sum_{j=1}^{2} \int d^{3} \vec{r}   \hat \psi_{j}^{\dagger}(\vec{r},t)\left( -\dfrac{\hbar^{2} \nabla^{2}}{2 m}+V_{i} (\vec{r},t)  \right) \hat \psi_{j} (\vec{r},t)  + \int d^{3} \vec{r} \hat \psi_{2}^{\dagger} (\vec{r},t) \hbar \omega_{a} \hat \psi_{2} (\vec{r},t)+H_{L}
\nonumber \\&& -\dfrac{\hbar}{2}\left\lbrace  \Omega_{i} e^{i \omega_{i} t} \int d^{3} \vec{r} \hat \psi_{1} (\vec{r},t) e^{-i \vec{k_{i}.\vec{r}}} \hat \psi_{2}^{\dagger} (\vec{r},t)+\sum_{\ell} \Omega_{d,\ell} e^{i \omega_{d,\ell} t} \int d^{3} \vec{r} \hat \psi_{2}^{\dagger} (\vec{r},t) e^{-i \vec{k_{d,\ell}.\vec{r}}} \hat \psi_{1} (\vec{r},t)\right\rbrace + c.c \nonumber \\&& + U \int d^{3} \vec{r} \hat \psi_{1}^{\dagger} (\vec{r},t) \hat \psi_{1}^{\dagger} (\vec{r},t) \hat \psi_{1} (\vec{r},t) \hat \psi_{1} (\vec{r},t).
\end{eqnarray}

Here, $H_{L}$ is the Hamiltonian for the lasers, $\omega_{a}$ is the atomic transition frequency and $\mu$ is the chemical potential. $\Omega_{i}$ and $\Omega_{d,\ell}$ are the Rabi frequencies of the incident and the $\ell$th order diffracted laser beams. The $\ell$th order Rabi frequency is related to the incident Rabi frequency as $\Omega_{d,\ell}$=$\Omega_{i}$ $exp ({i \ell \pi/2})$ $J_{\ell} (\epsilon k_{L} L/2 \sqrt{n^2-\sin^{2}{\theta_{i}}})$ \cite{Born}, $J_{\ell}$ is the $\ell$th order Bessel function , $n$ is the refractive index and $L=10 \mu m$ is the thickness of the condensate. $U=\dfrac{4 \pi \hbar^{2} N a_{s}}{m}$ is the two body interaction parameter, where $a_{s}$ is the interatomic $s$-wave scattering length (taken to be positive) and $m$ is the mass of the atom. 

 We assume that the population in the excited state is extremely small so that the atom-atom interaction in the excited state is neglected. Here, $\hat \psi_{1}$ and $\hat \psi_{2}$ are the ground and excited state field operators respectively. $k_{i}$($\omega_{i}$) and  $k_{d}$($\omega_{d}$)is the wavenumber (frequency) of the incident and diffracted light respectively. $V_{i} (i=1,2)$ is the trap potential of the ground state ($i=1$) and the excited state ($i=2$).

We can now write down the Heisenberg equation of motion for atomic field operators, $\hat \psi_{1}$ and $\hat \psi_{2}$ and then eliminate the excited state under the condition that the detuning $\delta=\omega_{i,d}-\omega_{a}$ is much larger than the natural line width of the atomic transition. The kinetic energy and the trap potential of the excited state is dropped under the assumption that the lifetime of the excited atom, which is of the order $1/\delta$, is so small that atomic center of mass motion may be safely neglected during this period. This yields the following equation of motion (the Gross-Pitaevskii equation) for the ground state in the mean field regime where we replace the atomic field operators by the corresponding $c$ numbers:

\begin{equation}
i \hbar \dfrac{\partial \psi_{1}}{\partial t}=\left[ -\dfrac{\hbar^2 \nabla^{2}}{2 m}+V_{1}(\vec{r},t)+V_{op}(\vec{r},t)-\mu \right]\psi_{1}+U \psi_{1}|^{2} \psi_{1} 
\label{mattereqn}
\end{equation}

\begin{equation}
V_{op}(\vec{r},t)=\dfrac{\hbar}{4 \delta}\left\lbrace |\Omega_{i}|^2+\sum_{\ell}|\Omega_{d,\ell}|^{2}+\sum_{\ell} \Omega_{i}^{*}\Omega_{d,\ell} e^{i \Delta \omega_{\ell} t} e^{-i \vec{\Delta k_{\ell}}.\vec{r}}+ \Omega_{d,\ell}^{*}\Omega_{i} e^{-i \Delta \omega_{\ell} t} e^{i \vec{\Delta k_{\ell}}.\vec{r}}\right\rbrace 
\label{optpot}
\end{equation}

Where, $\Delta \omega_{\ell}=\omega_{d,\ell}-\omega_{i}= \ell \omega_{s}$ and $\Delta \vec{k_{\ell}}=\vec{k}_{d,\ell}-\vec{k}_{i}= \ell \vec{k}_{s}$ are given by the Bragg diffraction condition \ref{sign1}, \ref{sign2}. Here $V_{1}=m/2(\Omega_{\bot}^{2}(t)(x^2+y^2)+\Omega_{||}^{2}z^2)$.  The first two terms in Eqn.(\ref{optpot}) is simply the static optical potential while the next two terms are the optical potential formed due to the interference between the $z$ component of the incident and diffracted light. The self consistent optical potential thus formed has the same spatiotemporal variation as the Faraday wave.  Actually, if the decay rate of the light field is much more than the parameter $\Omega_{i} \Omega_{d, \ell}/\delta$ then the optical potential formed will have a negligible influence on the acousto-optic effect. The chemical potential $\mu$ in the Thomas-Fermi approximation is determined by the number of particles, the scattering length $a_{s}$ and the trap parameters as \cite{Pethick}

\begin{equation}
\mu=\dfrac{15^{2/5}}{2}\left( \dfrac{Na_{s}}{a_{mean}}\right)^{2/5}\hbar \bar \omega .
\end{equation}

Where $\bar \omega = (\omega_{x} \omega_{y} \omega_{z})^{1/3}$, $ a_{mean} = \sqrt{\hbar/m \bar \omega}$ , $\omega_{x}$,$\omega_{y}$ and $\omega_{z}$ are the trap frequencies along the $x,y,z$ directions respectively and $a_{s}$ is the scattering length for Rubidium atoms. All lengths are scaled with respect to $a_{\bot}=\sqrt{\hbar/m \bar \Omega_{\bot}}$ and all frequencies are scaled with respect to $\bar \Omega_{\bot}$, where $\bar \Omega_{\bot}$ is the amplitude of the trap frequency along the radial direction and $m$ is the mass of the atom. The spatiotemporal oscillations of the BEC density along the weakly confined space occur at half the trap modulation frequency ($\omega_{s}= \Omega/2$). Here we are neglecting the damping of the Faraday waves. This can be approximately true if we have a strong and constant driving so that a steady Faraday pattern is maintained during the process of diffraction. Inclusion of damping only changes the threshold for pattern formation. We take the experimental conditions described in Ref. \cite{Engels07}. Specifically, we take the number of condensed $^{87} Rb$ atoms as $N=5 \times 10^5$ contained in a magnetic trap with frequencies $\left\lbrace {\bar \Omega_{\bot}/(2 \pi),\bar \Omega_{||}/(2 \pi)}\right\rbrace = \left\lbrace {160.5,7}\right\rbrace  Hz$ where the radial trap frequency $\Omega_{\bot}(t)=\bar \Omega_{\bot}(1+\alpha \cos{\Omega t})$ has a modulation of $20 $ percent ($\alpha=0.2$) and a modulation frequency $\Omega/ \bar \Omega_{\bot}=\omega=2$. In terms of the scaled variables, Eqn.(6) is reduced to a dimensionless one dimensional equation (this can be done by separating the wavefunction into a radial and a longitudinal part \cite{Stalinus04}). The radial part is approximately taken to be Gaussian and the radial part is integrated out and we are left with an effectively one dimensional equation:

\begin{equation}
i  \dfrac{\partial \phi(Z)}{\partial \tau}=\dfrac{1}{2}\left[ -\dfrac{ \partial^{2}}{\partial Z^2}+\omega_{Z}^{2}Z^2+V_{op}(Z,\tau)-\tilde \mu \right]\phi(Z)+ \dfrac{\omega_{\bot}}{2}|\phi(Z)|^{2} \phi(Z) 
\label{mattereqn2}
\end{equation}

\begin{equation}
V_{op}(Z, \tau)=\dfrac{1}{2 \delta \bar \Omega_{\bot}} \left\lbrace |\Omega_{i}|^{2}+\sum_{\ell}|\Omega_{d,\ell}|^{2}+\sum_{\ell} \Omega_{i}^{*}\Omega_{d,\ell} e^{i \tilde \omega_{s} \tau} e^{-i \tilde k_{d} Z} + \Omega_{d,\ell}^{*}\Omega_{i} e^{-i \tilde \omega_{s} \tau} e^{i \tilde k_{d} Z} \right\rbrace 
\label{optpot2}
\end{equation}

Here $\tilde \mu = 2(\bar \mu-\omega_{\bot})$ and the dimensionless parameters are defined as : $\bar \mu=\mu/\bar \Omega_{\bot}$, $\omega_{\bot}=\Omega_{\bot}/\bar \Omega_{\bot}$, $\tilde \omega_{s}=\omega_{s}/ \bar \Omega_{\bot}$, $\omega_{Z}=\Omega_{||}/\bar \Omega_{\bot}$, $\tau=\bar \Omega_{\bot} t$, $Z=z/a_{\bot}$ and $\tilde k_{d}/a_{\bot}$. The above equations are solved numerically and the resulting normalized deviation from initial density (scaled central density) is plotted in Fig. 2 for $\ell=0, \pm 1$, $\tilde \mu =2$, $\omega=2$, $\alpha=0.2$ and for two values of the strength of the optical potential $|\Omega^{0}|^{2}/2 \delta \bar \Omega_{\perp}=0$(thick line) and $|\Omega^{0}|^{2}/2 \delta \bar \Omega_{\perp}=0.8$(thin line). Here we have taken $\Omega_{i}$ and $\Omega_{d,l}$ as real and $\Omega_{i}=\sqrt{3} \Omega_{d,\ell}=\Omega^{0}$. The solutions are found to be relatively more unstable in the presence of the light field since in the presence of the light field, the amplitude of the perturbation grows much faster. The faster growth of perturbations in the presence of the light field is probably due to excitation of the several other modes which makes the system unstable. Actually for a constant driving the condensate behaves like an impact oscillator which is accompanied by increasing transverse cloud radius with time \cite{Engels07}.

\begin{figure}[t]
\hspace{-1.5cm}
\includegraphics{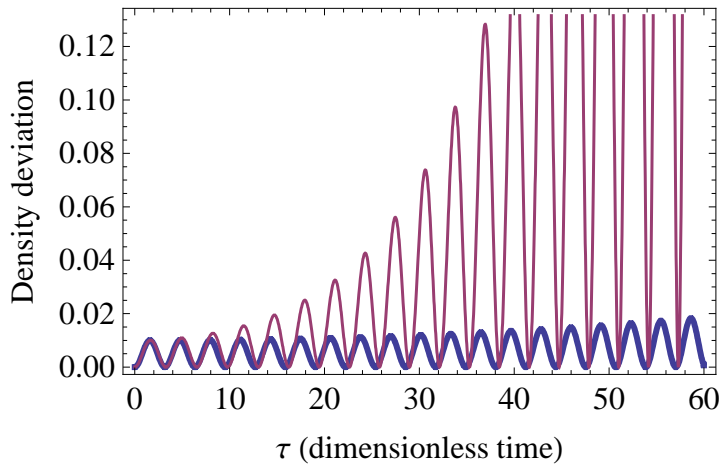} 
\caption{Plot of deviation from initial density for  $\ell=0, \pm 1$, $\tilde \mu =2$, $\omega=2$, $\alpha=0.2$ and for two values of the strength of the optical potential $|\Omega^{0}|^{2}/2 \delta \bar \Omega_{\perp}=0$(thick line) and $|\Omega^{0}|^{2}/2 \delta \bar \Omega_{\perp}=0.8$(thin line). The solutions are found to be relatively more unstable in the presence of the light field.}
\label{figure2}
\end{figure}

\section{Two-Level Model of the Quantum Acousto-Optic Effect}

In this section, we adopt a semi-classical approach to investigate the problem of Bragg scattering of light from Faraday waves travelling along the long axis of an elongated cigar shaped BEC. The Faraday pattern consists of Faraday waves travelling in opposite direction with wavevectors ${\vec{k_{s}}, -\vec{k_{s}}}$. We consider the situation as shown in Fig. 1 where an incident photon of frequency $\omega_{i}$ and wavevector $k_{i}$ is scattered by a Faraday wave phonon of frequency $\omega_{s}$ and wavevector $k_{s}$ travelling in the positive $z$ direction. The energy and momentum conservation condition requires that the diffracted photon have a frequency $\omega_{i}+\omega_{s}$ and wavenumber $k_{i}+k_{s}$.  The interaction in the case of scattering of incident photons by Faraday wave travelling in the negative $z$ direction can be viewed in the following terms. A photon with energy $\hbar \omega_{i}$ and momentum $\hbar \vec{k}_{i}$ is incident on a Faraday wave of frequency $\omega_{s}$ and momentum $-\hbar \vec{k}_{s}$. The incident photon is annihilated, giving rise instead to a new photon at $\omega_{d}$, $\vec{k}_{d}$ and a phonon $\omega_{s}$, $\vec{k}_{s}$ travelling in the positive $z$ direction. The above two interactions effectively involves only three momentum levels i.e. level with zero momentum and that with $\pm \hbar \vec{k}_{s}$. The loss due to decoherence and decay of the light field is taken into account. The above process is described by the quantized Hamiltonian:

\begin{eqnarray}
H&=&\sum_{j=1}^{2} \int d^{3} \vec{r} \hat \psi_{j}^{\dagger}(\vec{r},t)\left( -\dfrac{\hbar^{2} \nabla^{2}}{2 m}+V_{i} (\vec{r},t)  \right) \hat \psi_{j} (\vec{r},t)  + \int d^{3} \vec{r} \hat \psi_{2}^{\dagger} (\vec{r},t) \hbar \omega_{a} \hat \psi_{2} (\vec{r},t)+\hbar \omega_{1} \hat a_{1}^{\dagger} \hat a_{1}+\hbar \omega_{2} \hat a_{2}^{\dagger} \hat a_{2}
\nonumber \\&& -i \hbar \left\lbrace g_{1}  \hat a_{1} \int d^{3} \vec{r} \hat \psi_{1} (\vec{r},t) e^{i \vec{k_{i}.\vec{r}}} \hat \psi_{2}^{\dagger} (\vec{r},t)+g_{2} \hat a_{2} \int d^{3} \vec{r} \hat \psi_{2}^{\dagger} (\vec{r},t) e^{i \vec{k_{d}.\vec{r}}} \psi_{1} (\vec{r},t)\right\rbrace + H.c \nonumber \\&& + U \int d^{3} \vec{r} \hat \psi_{1}^{\dagger} (\vec{r},t)\hat \psi_{1}^{\dagger} (\vec{r},t) \hat \psi_{1} (\vec{r},t) \hat \psi_{1} (\vec{r},t)
\label{Ham2}
\end{eqnarray}

Here $\hat a_{1}$ and $\hat a_{2}$ are the destruction operators for the incident and the diffracted light respectively and $g_{1}$ and $g_{2}$ are the atom field electric-dipole coupling constant for the incident and the diffracted light fields respectively. The ground and excited state atomic field operators are $\hat \psi_{1}$ and $\hat \psi_{2}$ respectively.  We can write down the equation of motion for the operators $\hat \psi_{1}$, $\hat \psi_{2}$, $\hat a_{1}$ and $\hat a_{2}$ from the Hamiltonian \ref{Ham2}. 

\begin{equation}
i \hbar \dfrac{\partial \hat \psi_{1}}{\partial t}=\left( -\dfrac{\hbar^{2} \nabla^{2}}{2m}+V_{1}\right) \hat \psi_{1}+i \hbar g_{1} \hat a_{1}^{\dagger}e^{-i \vec{k}_{i}.\vec{r}} \hat \psi_{2}+i \hbar g_{2} \hat a_{2}^{\dagger} e^{-i \vec{k}_{d}.\vec{r}} \hat \psi_{2}+U | \hat \psi_{1}|^{2} \hat \psi_{1}. 
\end{equation}

\begin{equation}
i \hbar \dfrac{\partial \hat \psi_{2}}{\partial t}=\left( -\dfrac{\hbar^{2} \nabla^{2}}{2m}+V_{2}\right) \hat \psi_{2}-i \hbar g_{1} \hat a_{1} e^{i \vec{k}_{i}.\vec{r}} \hat \psi_{1}-i \hbar g_{2} \hat a_{2} e^{i \vec{k}_{d}.\vec{r}} \hat \psi_{1}+ \hbar \omega_{a} \hat \psi_{2}.
\end{equation}

\begin{equation}
i \hbar \dfrac{d \hat a_{1}}{d t}=\hbar \omega_{1} \hat a_{1}+i \hbar g_{1} \int d^{3}r \psi_{1}^{\dagger} e^{-i \vec{k}_{i}.\vec{r}} \psi_{2}-\kappa \hat a_{1}.
\end{equation}

\begin{equation}
i \hbar \dfrac{d \hat a_{2}}{d t}=\hbar \omega_{2} \hat a_{2}+i \hbar g_{2} \int d^{3}r \psi_{1}^{\dagger} e^{-i \vec{k}_{d}.\vec{r}} \psi_{2}-\kappa \hat a_{2}.
\end{equation}

Here, $\kappa \sim c/2L$ is the phenomological radiation loss term taken to be same for both the incident and the diffracted beam and $L$ is the length of the condensate \cite{Fallani}. Introducing the operators $\psi_{2}'= \psi_{2} e^{-i (\omega_{i}+\omega_{d})/2 t}$ and $\hat b_{j}=\hat a_{j} e^{-i \omega_{j} t}$, $j=1,2$, we adiabatically eliminate the excited state. We now replace the bosonic operators in the above equations by the coherent wavefunction of the condensate $<\hat \psi_{j}>=\psi_{j}$ (mean-field approximation)and the classical light fields amplitudes $<\hat b_{j}>=b_{j}$ (in the case of large intensity of the light fields). Integrating out the radial dimension, we get the following equations of motion for the new variables in one dimension:

\begin{equation}
i \hbar \dfrac{\partial \psi_{1}}{\partial t}=\left\lbrace -\dfrac{\hbar^{2} }{2m } \dfrac{\partial^{2}}{\partial z^{2}}-\mu_{1}+\dfrac{\hbar}{\delta}\left[ g_{1}^{2} b_{1}^{*} b_{1}+g_{2}^{2} b_{2}^{*} b_{2}+g_{1}g_{2} b_{1}^{*} b_{2}e^{-i \omega_{s} t} e^{i k_{s} z}+ g_{1}g_{2} b_{2}^{*} b_{1}e^{i \omega_{s} t} e^{-i k_{s} z}\right] \right\rbrace \psi_{1} +U(t)|\psi_{1}|^{2} \psi_{1},
\label{matter1} 
\end{equation}

\begin{equation}
\dfrac{d b_{1}}{dt}=-i\dfrac{g_{1}^{2}}{\delta}b_{1}-i G' b_{2} e^{-i\omega_{s} t} \int dz e^{i k_{s} z} |\psi_{1}|^{2}- \kappa b_{1},
\label{light1}
\end{equation}

\begin{equation}
\dfrac{d b_{2}}{dt}=-i\dfrac{g_{2}^{2}}{\delta}b_{2}-i G' b_{1} e^{i\omega_{s} t} \int dz e^{- i k_{s} z} |\psi_{1}|^{2}-\kappa b_{2},
\label{light2}
\end{equation}

\begin{figure}[t]

\begin{tabular}{cc}
×\includegraphics{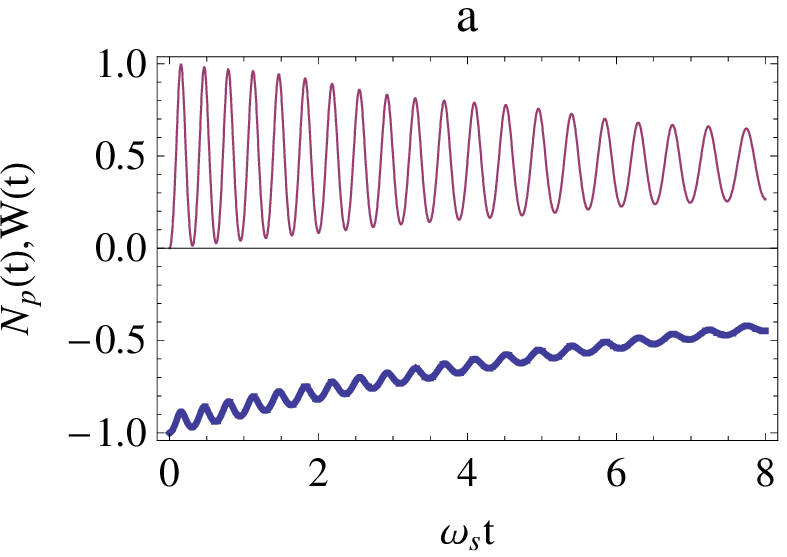}& \includegraphics{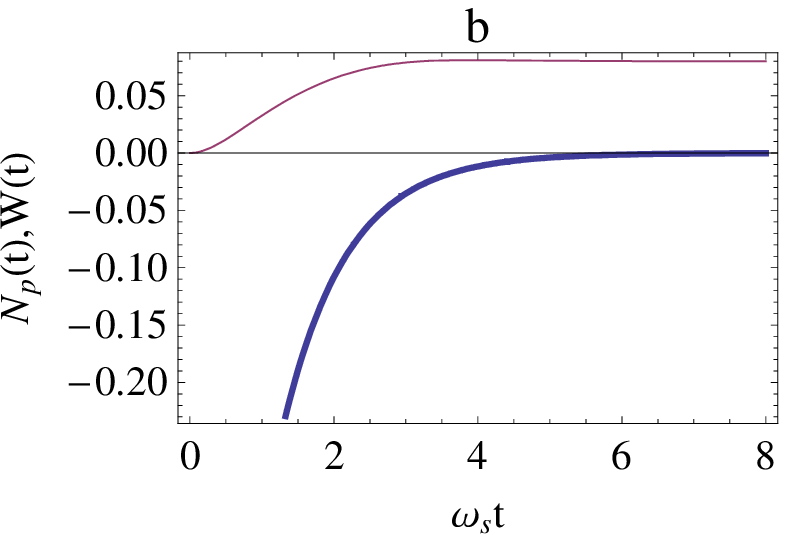}\\
 \includegraphics{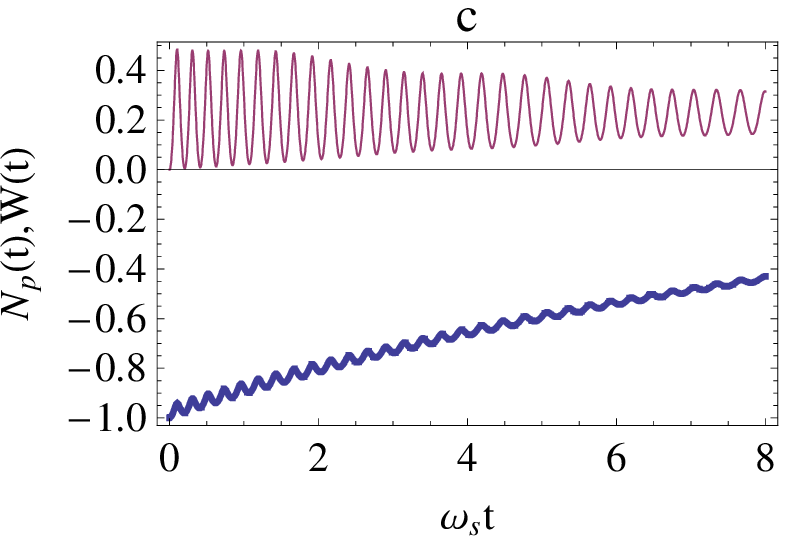} &\includegraphics{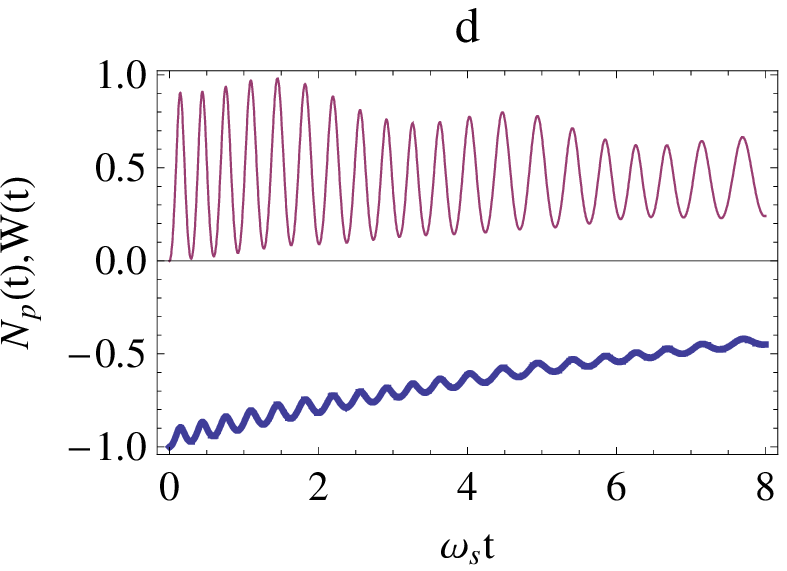} 
\end{tabular}

\caption{ Time evolution of the normalized population difference($W(t)$, thin curve) between the two momentum modes and the intensity difference ($N_{p}(t)$, thick curve) between the diffracted light beam and the incident beam. (a) The parameters chosen are: $g=10$, $g'=.1$, $\alpha=0.1$, $\kappa=0.1$ ,$\kappa'=0.1$, $\gamma=0.1$, $\omega=2$, $N=10^{5}$, $a_{s}/a_{mean}=0.00016$, $\dfrac{|\Omega^{0}|^2}{\delta \bar \Omega_{\perp}}=0.1$, $\dfrac{\bar \Omega_{||}}{\bar \Omega_{\perp}}=\dfrac{7}{160}$. The population in the two momentum modes undergo a damped Rabi oscillations due to Bragg scattering of the atoms from the optical grating when the light-atom coupling is greater than the loss rate of the light field. In this regime, the Rabi oscillations has a small contribution from the quantum acousto-optic effect and this is visible as a small beat frequency in the oscillations. The intensity difference, $N_{p}(t)$ also undergoes oscillations and approaches a steady value after some time.  (b) In the opposite limit when the decay rate of the light field ($\kappa=1.1$) is larger compared to the light-atom coupling ($g=1.0$), the Rabi oscillations are highly damped and the initial population difference decays to a stationary value. (c) On increasing the particle number $N=5\times10^{5}$, the amplitude of the Rabi oscillations decreases. (d) When the amplitude of the perturbation is increased, $\alpha=0.5$, the beat frequency in the Rabi oscillations becomes strong due to an increased contribution from the quantum acousto-optic effect.}

\label{figure5}
\end{figure}

Where, $G'=\dfrac{g_{1}g_{2}}{\delta} \pi \left( \dfrac{a_{\perp}}{4 \pi a_{s} N}\right)^{2/3} $, $\mu_{1}=\bar{\mu}-\hbar \Omega_{\perp}$ and $U(t)=U(1+\alpha \cos{\Omega t})$. The Eqns. \ref{matter1}, \ref{light1} and \ref{light2} are the coupled matter-wave and the light fields which determine the complete dynamics of the system. The third term on the right-hand side of Eqn. \ref{matter1} represents the optical potential (with the same periodicity as that of the Faraday wave) created by the interference between the $z$ component of the incident and the diffracted light fields and whose amplitude depends on time according to Eqns. \ref{light1} and \ref{light2}. This optical grating Bragg scatters the condensate atoms. The second terms on the right-hand side of Eqns. \ref{light1} and \ref{light2} are the matter-wave Faraday grating. An important point to note is that in deriving the Eqns.\ref{matter1}, \ref{light1} and \ref{light2}, we have assumed the Bragg condition \ref{Braggcond1}.  If the condensate ($\sim 125 \mu m$) is much larger than the Faraday wavelength($10 \mu m$) then periodic boundary condition can be assumed on $k_{s}z$ and the wavefunction can be written as a Fourier series \cite{Fallani}

\begin{equation}
\psi_{1}(k_{s}z)=\sum_{n} c_{n}(t) u_{n}(k_{s}z)e^{-in \omega_{s} t}.
\end{equation}

Where $u_{n}(k_{s}z)=\dfrac{1}{\sqrt{2 \pi}} exp(in k_{s}z)$ are the momentum eigenfunctions with eigenvalues $p_{z}=n(\hbar k_{s})$.  If we work close to the Bragg regime and the condensate is homogeneous along the long axis (this is true for a weak confinment along the long axis), we can assume that the only two momentum levels involved in the process are $n$ and $n+1$ (the two level approximation is true only if one diffracted beam is considered). We get the following equations for the variables $W=|c_{n}|^{2}-|c_{n+1}|^{2}$, $S=c_{n}c^*_{n+1}$, $A=i b_{1}b^{*}_{2}$, $N_{p}=b^{*}_{2}b_{2}-b^{*}_{1}b_{1}$ 

\begin{equation}
\dfrac{d W}{d \tau}=-2g\left(A^{*} S+A S^{*} \right) 
\end{equation}

\begin{equation}
\dfrac{d S}{d \tau}=-i\left(\Delta_{n}- U'(\tau) W \right)S+g A W-\gamma S 
\label{Bragg1}
\end{equation}

\begin{equation}
\dfrac{d A}{d \tau}=g' S N_{p}-\kappa A+i \kappa' A
\end{equation}

\begin{equation}
\dfrac{d N_{p}}{d \tau}=\dfrac{g'}{g}\dfrac{d W}{d \tau}-\kappa N_{p}
\end{equation}

In the above equations, all the physical quantities have been made dimensionless according to the scaling given in the previous section. Here, $g=\dfrac{g_{1}g_{2}}{\delta \bar \Omega_{\bot}}$, $g'=G'/\bar \Omega_{\bot}$,$\Delta_{n}= (\delta_{n}-\delta_{n+1})/\bar \Omega_{\bot}$, $\delta_{n}=\dfrac{n^2\hbar k_{s}^{2}}{2m}-n \omega_{s}$, $U'(\tau)= U(\tau)/\hbar \bar \Omega_{\bot}V$ ($V$ is the volume of the condensate),$\kappa'=\dfrac{g_{2}^2-g_{1}^2}{\delta \bar \Omega_{\bot}}$ and $\kappa\rightarrow \kappa/\bar \Omega_{\bot}$. We have introduced by hand $\gamma$ (dimensionless w.r.t $\bar \Omega_{\bot}$) as the decay rate of the atomic coherence between the two motional states $n$ and $n+1$. The decoherence is induced by spontaneous emission, phase diffusion, Doppler and inhomogeneous broadening \cite{Piovella, Fallani}. The dynamics of the system shows a phase matching between two kinds of phenomena, the acousto-optic effect and the Bragg scattering of atoms from the optical potential formed due to the interference of the incident and the diffracted light. The usual acousto-optic effect is dominant when either the atomic decoherence rate $\gamma$ or the radiation loss term $\kappa$ are greater than $g$ and $g'$ and the Bragg scattering of the atoms is dominant mechanism in the opposite limit. In the absence of interactions ($U=0$), $\Delta_{n}=0$ is the atomic Bragg scattering condition, arising from momentum and energy conservation. The two motional states assumed to be involved in the acousto-optic effect are the $n=0$ (stationary condensate state without any Faraday wave) and the $m=n+1=1$ (state with one unit of $\hbar k_{s}$).

The above equations are solved numerically in Fig.\ref{figure5}a for conditions $g=10$, $g'=.1$, $\alpha=0.1$, $\kappa=0.1$ ,$\kappa'=0.1$, $\gamma=0.1$, $\omega=2$, $N=10^{5}$, $a_{s}/a_{mean}=0.00016$, $\dfrac{|\Omega^{0}|^2}{\delta \bar \Omega_{\perp}}=0.5$, $\dfrac{\bar \Omega_{||}}{\bar \Omega_{\perp}}=\dfrac{7}{160}$. The initial conditions used are $W(0)=0$,$N_{p}(0)=-1$,$A(0)=0$, $S(0)=0.1$. We found that the conclusions that follow does not depend on the initial conditions, however a small but finite value of $S(0)$ is required to observe the dynamics (this implies that an initial coherence between the momentum modes is needed for the dynamics to start ). The diffracted light field grows in time and reaches a steady value and follows $W(t)$. As the diffracted field intensity grows, the population of the lower momentum state $|c_{n}|^2$ also grows since the condensate in the higher momentum state relaxes to the lower level after transferring momentum to the light field. After some time due to the Bragg scattering, the atoms perform a weakly damped Rabi oscillation caused by Bragg transition between the two momentum states $p=0$ and $p=\hbar k_{s}$. This process continues till a steady state is reached by both the atomic population difference ($W(t)$) and the intensity difference $N_{p}(t)$, due to a finite decay of the atomic coherence and the light field. Note that till time $t=60 ms$, only about less than half the energy is transferred to the diffracted beam due to the Bragg scattering of atoms. The rate at which atoms are transferred between the two momentum modes due to the Bragg scattering is different from that due to the acousto-optic effect and this results in the presence of beats in the Rabi oscillations in Fig.\ref{figure5}a.

The Rabi oscillations are suppressed if the decoherence factor $\gamma$ or the light field decay rate $\kappa$ is greater than $g$. This is illustrated in Fig.\ref{figure5}b with $g=1$ and $\kappa=1.2$. In this regime, the initial population difference decays to a stationary value without undergoing any significant Rabi oscillations and the acousto-optic effect dominates the Bragg scattering effect. In this regime, we observed that since the acousto-optic effect is the dominating process, half the incident light energy is transferred to the diffracted beam in a short time of about $t=30 ms$. There are two additional parameters, the time dependent interaction $U(\tau)$ and strength of the perturbation $\alpha$ which can influence the dynamics of the system. On increasing the interaction parameter by increasing the number of particles, the Bragg scattering mechanism becomes less efficient and the amplitude of the Rabi oscillations are reduced (as illustrated in Fig.\ref{figure5}c), probably due to a dephasing caused by an increased detuning from the Bragg resonance condition. As seen from Eqn.\ref{Bragg1}, the interaction term has a dynamical dispersive effect on the Bragg resonance. An increased number of particles also means that the wavenumber $k_{s}$ of the Faraday pattern decreases, i.e loss of the Faraday pattern. As a result the acousto-optic effect becomes less efficient and is not able to tranfer energy and momentum from the phonons  to the light field. This also contributes to the observed decreased amplitude of the Rabi oscillations in Fig.\ref{figure5}c and hence a decreased transfer of energy to the diffracted beam per Rabi cycle. The effect of increasing the strength of the perturbation is shown in Fig.\ref{figure5}d. The beats in the Rabi oscillations are found to be enhanced, due to an increased contribution from the acousto-optic rate.

Finally we would like to mention that coherence time which is basically determined by the amount of time the diffracted photons stay inside the condensate should be high. This can be achieved by keeping the incident angle $\theta_{i}$ large so that the incident photon after diffraction travels a longer path inside the condensate before it leaves the condensate.

\section{Conclusion}

In conclusion, we have studied light scattering in the Bragg regime from an elongated cigar shaped BEC in the presence of a sound-like wave (Faraday waves) along the long axis. As in the case of the classical acousto-optic effect we show that the Faraday waves are able to transfer energy to the light field. We find that under appropriate conditions, there are two phase matching phenomena, the quantum acousto-optic effect (Bragg scattering of light by the dynamic matter-wave grating due to the Faraday waves) and the Bragg scattering of the BEC atoms from the optical grating formed in the BEC due to the interference between the incident and the diffracted laser beams. In section II, we have shown the system is relatively more unstable in the presence of the light field. In section III, we showed using the Heisenberg equation of motion for the atomic and light field amplitudes, that by varying the intensity of the incident light, one can switch between the acousto-optic regime and the atomic Bragg scattering regime. In particular, when the atom-field interaction is more that the light field decay rate, the atomic Bragg scattering dominates as seen by the Rabi oscillations between the two BEC momentum modes and in the opposite limit the acousto-optic effect dominates. On increasing the particle number, both the atomic Bragg scattering and the acousto-optic effect become less efficient as observed from the reduced Rabi oscillation and reduced amount of power transferred to the diffracted beam per Rabi cycle. 

\section{Acknowledgements}

The author thanks J.M. Rost and C. Ates for useful discussions.

\end{document}